\documentclass[12pt,twocolumn]{article}

\def\ii{\'{\i }}
\usepackage{graphics}
\usepackage{epsfig}
\usepackage{amssymb}
\begin{document}
\title{Strong color fields and heavy flavor production}
\author{I. Bautista and C. Pajares\footnote{IGFAE and Departamento
de F\ii sica de Part\ii culas, Univ. of Santiago de Compostela,
15706, Santiago de Compostela, Spain}}
\maketitle
\begin{abstract}
The clustering of color sources provides a natural framework for soft
partonic interactions producing  strong color fields. We study the
consequences of these color fields in the production of heavy flavor and the behavior of the
nuclear modification
factor.
\end{abstract}
\section{Introduction}
Heavy flavor production in heavy ions collisions is an ideal
probe to study the early time dynamics of these nuclear
collisions. Several theoretical studies predict
\cite{Muller:1992xn} a substantial enhancement of open charm
production associated to deconfined parton matter relative to the
case of a purely hadronic scenario without quark-gluon plasma
formation. Recent studies point out that the dynamics of heavy
quarks is dominated by partonic interactions in a strongly coupled
plasma modeled neither by hadronic interactions nor by
color screening alone \cite{Linnyk:2008hp}. Therefore, these
quarks are very relevant in the
study of the initial state of the collision. Owing to difficulties in 
reconstruct the $D$-mesons decay
vertex, RHIC experiments have measured open charmed quarks indirectly, via
the semileptonic decay to nonphotonic electrons or muons
\cite{Adams:2004fc} \cite{Adler:2004ta}. In the standard picture
charm quarks are produced by initial gluon fusion and their
production rates are expected to be well described by perturbative
quantum chromodynamics (pQCD) at fixed order plus next-to-leading
logarithms (FONLL) \cite{Cacciari:1998it}. The suppression of 
single, nonphotonic electrons or muons at RHIC is usually
attributed to heavy-quark energy loss. As a charmed quark of
energy $E$ cannot radiate gluons forming an angle below $\arcsin
(m/E)$ (dead cone effect), it is expected that heavy quarks lose 
less energy than light quarks \cite{Dokshitzer:2001zm}, but the
suppression experimentally observed is similar. In fact, many
calculations based on  energy loss via hard scattering
\cite{Djordjevic:2005db} or via multiple soft
 collisions \cite{Armesto:2005mz} obtained less suppression than the
experimental data when the beauty contribution is taken into
account. Similar results are obtained in evaluations based on
medium interactions or collisional dissociation
\cite{Adil:2006ra}. However, it has been argued
\cite{MartinezGarcia:2007hf}\cite{Zhou:2009zzg}
 that under the assumption
of an enhancement of the heavy-quark baryon-to-meson ratio, analogous to
the
case of the proton-to-pion and the $\Lambda$-to-kaon ratios measured in
Au-Au
collisions at RHIC, it is possible to achieve a larger suppression of the
nuclear modification factor.
This is possible, because the heavy-quark mesons have a larger branching
ratio to decay inclusively into electrons as compared to heavy-quark
baryons
and therefore, when the former are less copiously produced in a heavy-ion
environment, the nuclear modified factor decreases, even 
in the absence of
heavy-quark energy loss. Indeed the single nonphotonic nuclear modified
factor, $R_{AA}^{e}$ can
be expressed as \cite{Ayala:2009pe}
$R_{AA}^{e}=R_{AA}^{D+\Lambda}F$ where $R_{AA}^{D+\Lambda}$ is the
nuclear modified factor for D and $\Lambda_{c}$, i.e.
\begin{equation}
R_{AA}^{D+\Lambda}=\frac{N_{AA}^{D}+N_{AA}^{\Lambda}}{N_{coll}
(N_{pp}^{D}+N_{pp}^{\Lambda})}
\end{equation}
with $N^{D}$ and $N^{\Lambda}$ the produced $D$ and $\Lambda$ in $AA$ or 
$pp$
collisions and the $N_{coll}$ is the number of collisions at a given
centrality. The factor $F$ is given by the expression
\begin{equation}
F=\frac{(1+a)(1+xCa)}{(1+Ca)(1+xa)}
\end{equation}
where, $a$ and $Ca$ are the charmed baryon-to-meson ratios in 
proton-proton
and $AA$ collisions respectively. Therefore $C$ represents the
enhancement factor for the ratio of charm baryons to mesons in $AA$
as compared to $pp$ collisions, $x$ is the ratio between the branching
ratios for the inclusive decay of $\Lambda$ and $D$ into electrons:
\begin{equation}
a=(\frac{\Lambda}{D})_{pp}\mbox{,   } Ca=(\frac{\Lambda}{D})_{AA}\mbox{,
} x=\frac{B^{\Lambda \rightarrow e}}{B^{D \rightarrow e}}
\end{equation}
In \cite{Ayala:2009pe} $x$ has been estimated to be 0.14. As long as $C$ 
is
larger than 1
the factor $F$ becomes less than 1 and $R_{AA}^{e}<R_{AA}^{D+\Lambda}$.
The
main question to solve is whether the expected charmed baryon-to-meson
expected enhancement is large enough to explain the difference with
the
experimental data.

 In a high-energy heavy-ion collision, strong color fields are expected
to be produced between the partons of the projectile and target
\cite{ToporPop:2007hb}\cite{Pop:2009sd}
\cite{Armesto:1994yg}. These color
fields are similar to those that appear in the glasma
\cite{Kharzeev:2003sk} produced
in the color glass condensate (CGC). In a string heavy-quark pairs are 
produced via the
Schwinger
mechanism with a rate $\Gamma_{Q \bar{Q}}=\exp{\left[-\frac{\pi
m_{Q}^{2}}{k}\right]}$
where $k$, is the effective string tension, proportional to the strength
of
the field (for a single string $k_{1}\sim$ 1 GeV/fm). Longitudinal string
 models predict for heavy flavor a very suppressed production rate,
since
\begin{equation}
\frac{\Gamma_{Q\bar{Q}}}{\Gamma_{q
\bar{q}}}=\exp{(\frac{-\pi}{k_{1}}(m_{Q}^{2}-m_{q}^{2}))}\ll 1
\end{equation}
for $Q=c$ and $q=u, d$. The color in these strings is confined to
a small area in the transverse space, $\pi r_{0}^{2}$, with
$r_{0}\simeq 0.25$fm. In a central heavy-ion collision many
strings are formed between the partons of the projectile and
target in a
limited collision area, starting to overlap each other,
forming clusters. The field strength of the cluster is
proportional to the square root of the number of strings. So, for
a cluster of nine strings, the string tension increases more than
eight orders of magnitude becoming comparable to the initial
FONLLpQCD. The effect of strong color fields for open charm has
been investigated before \cite{ToporPop:2007hb} showing that a
three- fold increase of the effective string tension results in a
sizable enhancement of the total charm cross section and the
nuclear modified factor shows a suppression at moderate $p_{T}$
consistent with the RHIC data.

In this paper, we study the effects of strong color fields in the
framework of percolation of strings \cite{Armesto:1996kt}. In this
framework, a strong color field is obtained inside the clusters
formed by the overlapping of individual strings. The clusters
behave like individual strings with a higher string tension owing to
the higher color field, and their energy momentum is the sum of
the energy-momenta of individual strings. The color field of the
string is stretched between a quark and an antiquark or between a
diquark and a quark located at the extremes of the string. In the
case of a cluster, instead of quarks or antiquarks we have
complexes $Q$ and $\bar{Q}$ formed from the different quarks and
their antiquarks or diquarks and quarks at the extremes of the
individual strings \cite{Amelin:1994mc}\cite{Amelin:1994uy}. The
clusters behave like strings with complexes $Q\bar{Q}$, located at
the end, decaying into new pairs $Q\bar{Q}$, $\bar{Q}Q$, until
they come to objects with mass comparable to hadron masses, which
are identified with the known hadrons by combining the produced
quarks or antiquarks with the appropriate statistical weights. In
this way, the production of baryons and antibaryons is enhanced
with the number of strings in the cluster. The cluster not
only has a stronger color field than the individual string giving rise
to a mass-enhancement effect but also enhances the production of
baryons relative to mesons owing to the increasing probability of
getting three quarks or three antiquarks from the complex
$Q\bar{Q}$ \cite{Amelin:1994mc}. This second effect is similar to
what happens in coalescence or recombination models
\cite{Hwa:2001ih}\cite{Fries:2003vb}.

The percolation of strings incorporates to some extent the
recombination of flavors in a dynamical way. Indeed a dynamical
quark recombination model has shown a sizable suppression factor
for the nonphotonic electrons nuclear modification factor
\cite{Ayala:2009pe}.

We evaluate the nuclear modification factor for
$D_{0}$, $\Lambda_{c}$ and B at RHIC energies, computing also the
baryon-to-meson ratio in $AA$ and $pp$ collisions. We observe also in
$pp$ a moderate enhancement of the ratio as a function of the
transverse momentum which has consequences concerning the value of
$F$ and therefore the rate of the nonphotonic electron
suppression. The plan of the paper is as follows: In the next
section, we introduce briefly the percolation of the strings, and
then we present our results and conclusions.

\section{The string percolation model}
In the string percolation model
\cite{Armesto:1996kt}\cite{DiasdeDeus:2003ei}\cite{Pajares:2005kk}
\cite{Braun:2000hd} \cite{Cunqueiro:2007fn}, multiparticle production is
described in terms of
color strings
stretched between the partons of the projectile and the target.
With increasing energy and/or atomic number of the colliding
particles, the number of strings $N_{s}$, grows and they start to
overlap forming clusters, very much like disks in two-dimensional
percolation theory. At a certain critical density, a macroscopical
cluster appears, which marks the percolation phase transition.
This density corresponds to the value $\eta_{c}=1.2-1.5$ (depending on
the profile function of the colliding nuclei) where
$\eta=N_{s}S_{1}/S_{A}$ and $S_{A} $ stands for the overlapping
area of the colliding objects. A cluster of $n$ strings behaves like a
single string with the energy momentum corresponding to the sum of
 individual ones and with a higher color field corresponding to
the vectorial sum in color space of the color fields of the
individual strings. In this way, the mean multiplicity $<\mu_{n}>$
and the mean transverse momentum squared $<p_{T n}^{2}>$ of the
particles produced by a cluster are given by
\begin{equation}
<\mu_{n}>=\sqrt{\frac{n S_{n}}{S_{1}}}<\mu_{1}> 
\end{equation}
and $<p_{Tn}^{2}>=\sqrt{\frac{nS_{1}}{S_{n}}}<p_{T1}^{2}>$,
where $<\mu_{1}>$ and $<p_{T 1}>$ are the corresponding quantities in a
single string.

In the limit of high density of strings, equations (5) transforms
into \cite{Braun:2000hd}
\begin{equation}
<\mu>=N_{s}F(\eta)<\mu_{1}>
\end{equation},
$<p_{T}^{2}>=\frac{<p_{T 1}^{2}>}{F(\eta)}$
 with $ F(\eta)=\sqrt{\frac{1-e^{-\eta}}{\eta}}$.

For a specific kind of particle $i$, we will use: $<\mu_{1}>_{i}$,
$<p_{T 1}^{2}>_{i}$, $<\mu_{n}>_{i}$, and $<p_{T n}^{2}>_{i}$ for
the corresponding quantities. To compute the
multiplicities, we must know $N_{s}$ and $\mu_{1}$ (so for a fixed
centrality, knowing $N_{s}$ we deduce the density $\eta$). Up to
RHIC energies, in the central rapidity region $N_{s}$ is
approximately twice the number of collisions, $N_{coll}$. However
$N_{s}$ is larger than $2N_{coll}$ at RHIC and LHC energies, in
the same way as in nucleon-nucleon collisions. According to color
exchange models, such as  dual parton model or the quark gluon string
model \cite{Capella:1978ig}\cite{Kaidalov:1982xe},
 the number of produced strings $N_{s}$ is larger
than two, starting at the RHIC energies. Indeed, at high enough
energy the strings are stretched not only between the
diquarks(quarks) and quarks(diquarks) of the projectile and target
respectively, but also between
quarks(antiquarks) and antiquarks(quarks) of the sea. As the
energy increases, more $q-\bar{q}$ or $\bar{q}-q$ are formed and
$N_{s}$ becomes larger than two. For the same reason in $AA$
collisions, $N_{s}$ at high energy is larger than $2N_{coll}$. In
this work we take the values of $N_{s}$ from a Monte-Carlo based
on the quark-gluon string model \cite{Amelin:2001sk}.

Concerning the transverse momentum distribution, one needs the
distribution
 $g(x,p_{T})$ for each cluster, and the
mean square transverse momentum distribution of the clusters
$W(x)$, where $x$ is the inverse of the mean of the squared
transverse momentum of each cluster which is related to the
cluster size by equation (5). We take
$g(x,p_{T}^{2})=\exp{(-p_{T}^{2}x)}$ as it is used for fragmentation of
the Lund string. For the weight function $W(x)$ we
take the gamma distribution. The generalized gamma distributions
are unique distributions stable under the cluster-size
transformations
\cite{DiasdeDeus:2003ei}\cite{JonaLasinio:1974rh}\cite{DiasdeDeus:1997di};
for simplicity we choose gamma
distribution
simplicity \cite{DiasdeDeus:2003ei}.
\begin{equation}
W(x)=\frac{\gamma(\gamma x)^{k-1}}{\Gamma(k)}\exp{(-kx)}
\end{equation}
with
\begin{equation}
\gamma=k/<x>
\end{equation}
and
\begin{equation}
\frac{1}{k}=\frac{<x^{2}>-<x>^{2}}{<x>^{2}}
\end{equation}

The function $k$ measures the width of the distribution (7) and is
the inverse of the normalized dispersion of the transverse
momentum squared. The function $k$ depends on the density of
strings $\eta$.

The transverse momentum distribution $f(p_{T},y)$ of particle $i$ is
\begin{equation}
f(p_{T},y)=\frac{dN}{dp_{T}^{2}dy}=\int_{0}^{\infty}dxW(x)g(p_{T},x)=
\end{equation}
\begin{center}
$\frac{dN}{dy}\frac{k-1}{k}F(\eta)\frac{1}{(1+F(\eta)
p_{T}^{2}/k<p_{T}^{2}>_{1i})^{k}}$
\end{center}

The formula (10) is valid for all types of collisions,
energies and also all kind of flavors.
 Later we will extend (10) for baryons.
The function $k(\eta)$ was determined by comparing (10) to RHIC data.
The function $k$
decreases with $\eta$ up to values $\eta\simeq 1$ (peripheral Au-Au collisions at RHIC energies)
 from there it increases slowly. This behavior was expected. In fact, at
low
 density there is no overlapping of strings and there are isolated 
strings;
therefore $k\rightarrow\infty$. When the density and therefore the
numerator of Eq. (9) increases then k decreases. The minimum of k
will be reached where the fluctuations in the cluster-size reach
its maximum. Above this point, increasing $\eta$ these
fluctuations decrease and $k$ increases. The agreement with data
for $p_{T}$ up to 5 GeV/c is very good
\cite{DiasdeDeus:2003ei}\cite{Pajares:2005kk}.

In percolation of strings the fragmentation of a cluster of many strings
 is via the Schwinger mechanism, producing successive pairs $Q\bar{Q}$,
 where $Q$ represents the complexes of quarks, diquarks and antiquarks
at the extremes of the original string. It is clear that formula (10)
only contains the effect of the stronger color field of the cluster,
 which enhances heavy particles production, irrespective of their being
 mesons or baryons,
but it does not contain the breaking via flavor complexes
 $Q\bar{Q}$ and therefore cannot describe baryons correctly.
 In previous papers
\cite{Amelin:1994mc}\cite{Amelin:1994uy}\cite{Amelin:2001sk}
Monte Carlo codes were presented where
 this mechanism was built up, but with the approximation of fusion
 of only two strings \cite{Armesto:1996kt}\cite{Amelin:1994mc} or using
an effective color field \cite{Amelin:2001sk}. To to keep a closed 
analytical formula, incorporating the
antibaryon
 and baryon enhancement from the mechanism depicted here, we observe
that this
enhancement is similar to using the formula (10) with a larger
density, or equivalently with a larger $N_{s}$. This means that
for antibayons or baryons if we want to continue with formula (10) we
must replace $\eta$ by $\bar{\eta}_{B}$,
\begin{equation}
\bar{\eta}_{\bar{B}}=N_{s}^{\alpha}\eta
\end{equation}
and instead of the first equation (6) we must use
\begin{equation}
\mu_{\bar{B}}=N_{s}^{1+\alpha}F(\eta_{B})\mu_{1\bar{B}}
\end{equation}
where the parameter $\alpha$ is obtained from a fit to the experimental
 dependence of the $p_{T}$-integrated $\bar{p}$ spectra with centrality
\cite{PHENIX}. The obtained value is $\alpha=0.09$. In same sense we can
say that the antibaryons (baryons)
probe a higher density than mesons for a fixed energy and type of
collisions.

The equations (10), (11) and (12) allow us to compute the antibaryon
(baryon) spectra.
 The equations (11) and (12) replace the recombination process described
in this section and it should be considered as an approximation to keep
the
analytical formula (10). The formulae (10), (11) and (12) are valid for
all kind of particles and not only for heavy flavor. We will show some
results
concerning light flavor.

\section{Results}
The equation (10) is limited to low and moderate $p_{T}$ not higher
than $4-5$ GeV/c.
 In fact, we consider a gaussian $p_{T}$ distribution for the particles produced
  from the fragmentation of a string, without any power-like tail. This excludes
  the high-$p_{T}$ behavior, although our formula (10) allows for an
interpolation from
  low to high $p_{T}$. By continuity, the high $p_{T}$ suppression observed at RHIC
  should give rise to a suppression at moderate $p_{T}$, say $4-5$ GeV/c,
which is
  the limit where our equations apply.

To know the $p_{T}$ distributions given by formula (6) we
need the values of $<p_{T}^{2}>_{1D}\simeq <p_{T}>^{2}_{1D}$ and
 $<p_{T}^{2}>_{1\Lambda_{c}}\simeq <p_{T}>^{2}_{1\Lambda_{c}}$, i. e.
the mean $p_{T}$ of $D$ and $\Lambda_{c}$ particles produced from
one string. We use $<p_{T}>_{1D}=1.5$ GeV/c and
$<p_{T}^{2}>_{1\Lambda_{c}}=1.9$GeV/c. The difference between
these two values is close to the difference between the masses of
$D_{0}$ and $\Lambda_{c}$ and also agrees with the difference between
the values commonly used of primordial transverse momentum of
pions and protons, $<p_{T}>_{1\pi}=0.2-0.3$ GeV/c,
$<p_{T}>_{1p}=0.6-0.7$ GeV/c. For B we use $<p_{T}>_{1B}=4.25$
GeV/c.

In formula (10) the normalization is established by the values of
$\frac{dN}{dy}$ at $p_{T}=0$, which are computed using the formulae
(6) for D and (9) for $\Lambda_{c}$. To do this, we use
the values
$\mu_{1D}=exp(-F(\eta)\frac{m_{D}^{2}}{<p_{T}>^{2}_{1D}})\mu_{1\pi}$
and
$\mu_{1\Lambda_{c}}=exp(-F(\eta_{\Lambda_{c}})m_{\Lambda_{c}}^{2}/<p_{T}>_{1\Lambda_{c}}^{2})\mu_{1\pi}$,
with $\mu_{1\pi}=0.8$ \cite{Cunqueiro:2007fn}. We use these
functions for $\mu_{1D}$ and $\mu_{\Lambda_{c}}$ because for heavy
particles, $m_{T}^{2}$ is very different from $p_{T}^{2}$.
Concerning the function $k(\eta)$, we take the shape and values
from the studies done in reference \cite{Cunqueiro:2007fn} for $AA$
collisions. In the case of $pp$ we take $k(\eta)=3.97$ at
$\sqrt{s}=200$ GeV and $k(\eta)=4.07 $ at $\sqrt{s}=5.5$ TeV. We
discuss later the sensibility of the obtained result for the ratio
$(\Lambda_{c}/D^{0})$ to different k values.

In fig.
1 we present our results for the nuclear modified factor $R_{AA}$
for Au-Au collisions at RHIC for $D^{0}$, $\Lambda_{c}$, 
$B$ and $R_{AA}^{e}$ using formulae (1) as a
function of
$p_{T}^{2}$ compared with
the experimental data on nonphotonic electrons.
\begin{figure}
  \includegraphics[height=.3\textheight]{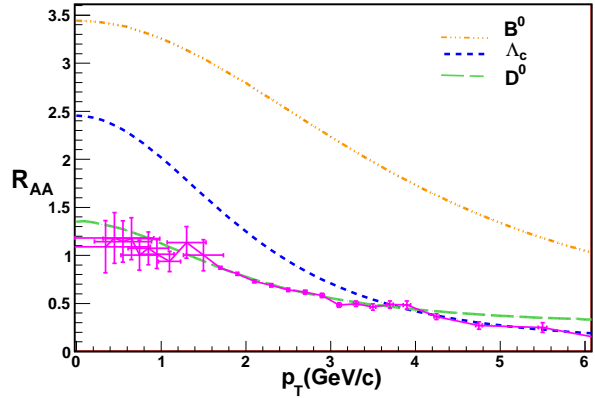}
  \caption{$R_{AA}$ for Au+Au  central collisions, bars is data taken from 
PHENIX \cite{PHENIX}.}
\end{figure}

The overall normalization is given by the value of $R_{AA}$ at
$p_{T}^{2}=0$ which has to do with the factor
$\exp{(F(\eta_{pp})-F(\eta_{AA}))\frac{m_{D}^{2}}{<p_{T}^{2}>_{1D}}}$.
Since we know the number of strings produced in $pp$ and $AA$ collisions,
we know $\eta_{pp}$, $\eta_{AA}$, $F(\eta_{AA})$ and $F(\eta_{pp})$
and the only free parameter is $<p_{T}^{2}>_{1D}$. From the data
we obtain $<p_{T}>_{1D}\sim 1.5$ GeV/c. The experimental errors
allow us a $15 \%$ freedom in the value of $<p_{T}>_{1D}$, however
a higher value than 1.5 GeV/c would not be realistic an a lower value will give rise to a higher normalization
and therefore $R_{AA}$ for $p_{T}> 4 $GeV/c will exceed the
experimental data even more than with the used value. The value of
$R_{AA}$ for $D^{0}$
at low $p_{T}$ agrees with the results in
\cite{Pop:2009sd}\cite{Ryskin:1999yq}, for $p_{T}>4$GeV/c we
obtain an $R_{AA}$ larger than the non-photonic leptonic data. In fig. 2,
we
present our results on $R_{AA}$ at $\sqrt{s}=5.5$ TeV for a
$D^{0}$, $\Lambda_{c}$ and B. We see that, as expected, as energy
increases the low $p_{T}$ $R_{AA}$ increases, although the
suppression at intermediate $p_{T}$ is similar.
\begin{figure}
  \includegraphics[height=.3\textheight]{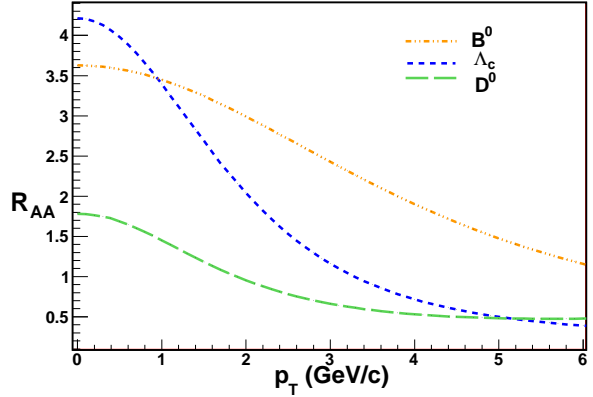}
  \caption{$R_{AA}$ for Pb+Pb  central collisions at $\sqrt{s}=5.5$ TeV.}
\end{figure}

In fig 3, we present the ratio $\Lambda_{c}/D^{0}$ for Au-Au at
$\sqrt{s}=200$GeV.  We observe  that the ratio increases up to a maximum
of 1.45 around $p_{T}\sim 4-5$ GeV/c. A very similar enhancement has been
obtained in the dynamical recombination model \cite{Ayala:2009pe}.
\begin{figure}
  \includegraphics[height=.3\textheight]{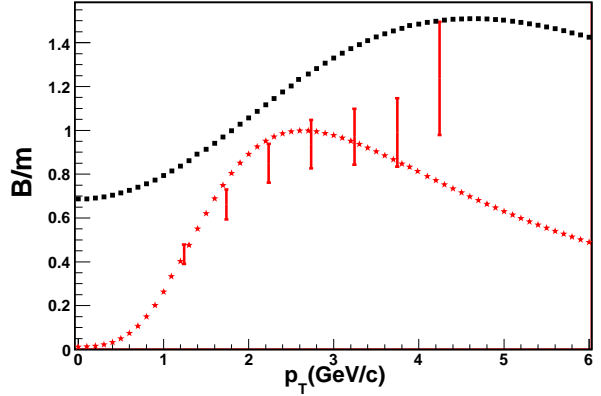}
  \caption{Squares are used for ratio $\Lambda_{c}/ D^{0}$, starts are 
used for $\bar{p}/\pi$, and errorbars are used for data from PHENIX for 
Au-Au  
central collisions at
$\sqrt{s}=200$ (GeV).}
\end{figure}

For comparison we include also our results for $\bar{p}/\pi$ at
central Au-Au collisions together with experimental data
\cite{Adler:2003cb}. In fig. 4 we
show the ratio $\Lambda_{c}/D^{0}$ for $pp$ at $\sqrt{s}=200$ GeV.
We observe a very smooth enhancement.
\begin{figure}
  \includegraphics[height=.3\textheight]{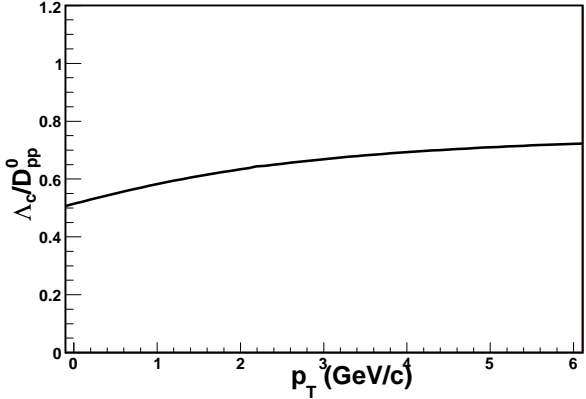}
  \caption{Ratio $\Lambda_{c}/ D^{0}$ for p+p collisions at
$\sqrt{s}=200$ GeV.}
\end{figure}
In fig 5 and fig. 6 we show our results for the ratio $\Lambda_{c}/D^{0}$
at
$\sqrt{s}=5.5$TeV for Pb-Pb collisions and $pp$ collisions respectively. 
We
observe in both of them larger enhancement than at RHIC energies,
particularly in the nuclear case.
\begin{figure}
  \includegraphics[height=.3\textheight]{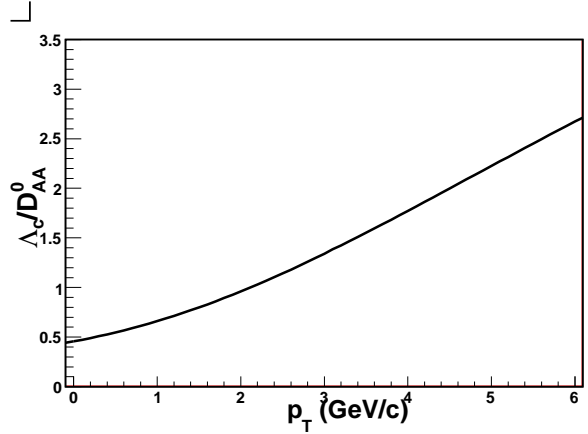}
  \caption{Ratio $\Lambda_{c}/ D^{0}$ for Pb-Pb central collisions at
$\sqrt{s}=5.5$ TeV.}
\end{figure}
\begin{figure}
  \includegraphics[height=.3\textheight]{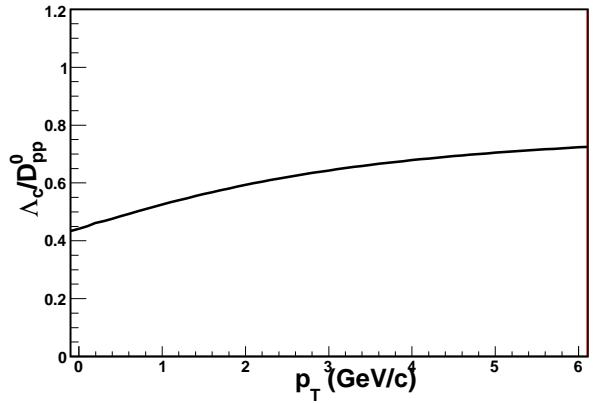}
  \caption{$R_{AA}$ $\Lambda_{c}/D^{0}$ for p+p  collisions at
$\sqrt{s}=5.5$ TeV.}
\end{figure}
\begin{figure}
  \includegraphics[height=.3\textheight]{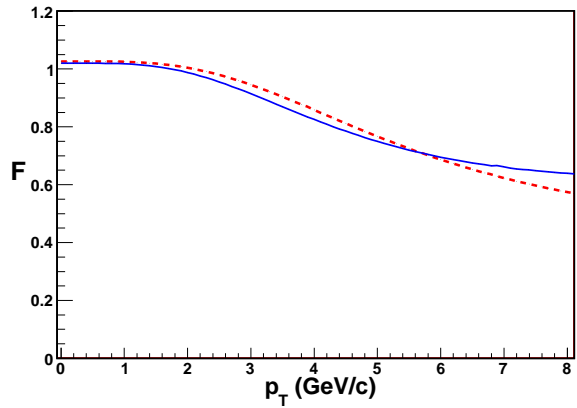}
  \caption{Factor F for  central collisions at RHIC in full line and LHC 
energies in dashed lines.}
\end{figure}
In fig. 7 we plot the factor F at $\sqrt{s}=200$ GeV (blue) and
at $\sqrt{s}=5.5$ TeV (red line). We observe that at RHIC energies
the factor $F$ is only slightly below one, and for $p_{T}\simeq
4-5$ GeV/c it is clearly over 0.5, which means that the
$\Lambda_{c}/D^{0}$ enhancement in $AA$ is not able to explain all
the difference between experimentally observed suppression of
$R_{AA}$ for non-photonic electrons and the pQCD
expectations. We have studied the effects due to the uncertainties
in the $k$ values for $pp$ at this energy. For reasonable
alternative $k$ values the enhancement of $\Lambda_{c}/D_{0}$ in pp
with $p_{T}$ is larger giving rise to lower C factor in
equation (3) and therefore the factor F is near to one, consistent with 
our main conclusion, namely that the $\Lambda_{c}/D^{0}$
enhancement is not able to explain all the difference between the
experimentally observed values and the perturbative expectations.

\section{Conclusion}
The overlapping of the strings formed in the collision of heavy
nuclei particles produces strong color fields which give rise to an
enhancement
of heavy flavor. We have computed the nuclear modification factor of
$D^{0}$, $\Lambda_{c}$ and $B^{0}$ at RHIC and LHC energies for
$AA$ collisions. Referring to $D^{0}$, we obtain a good agreement
at low $p_{T}$ with the experimental data for the nuclear modification
factor of non-photonic electrons. For $p_{T}$ values between 2 and 6 
GeV/c our results obtained are over
 the experimental data as pQCD \cite{Amelin:1994mc}.

The ratio $\Lambda_{c}/D^{0}$ as a function of $p_{T}$ for Au-Au
collisions at $\sqrt{s}=200$ GeV is enhanced showing a maximum
around 5 GeV/c. Such as enhancement is much larger at LHC
energies. However, the enhancement $\Lambda_{c}/D^{0}$ can explain
only half of factor 2 difference between the experimental data
and the pQCD expectations at RHIC energies.

In p-p collisions the ratio $\Lambda_{c}/D^{0}$ also rises as a
function of $p_{T}$ but very smoothly at RHIC energies. At LHC this
increase is a factor of 2 between $p_{T}=0$ and $p_{T}=6$ GeV/c.
The enhancements of $\Lambda_{c}/D^{0}$ in AA and pp collisions are larger
at LHC than at RHIC as it was expected due to the stronger
color fields produced.

\section{Acknowledgments} We thank N. Armesto and A. Ayala and M. A.
Braun for discussions. This research was supported by MICINN of Spain,
under Grant
FPA2008-01177 and by Xunta de Galicia.


\end{document}